\documentclass{ws-procs9x6}
\usepackage{graphics}
\begin{document}

\title{NEMO 3 experiment: preliminary results from 2003 data}

\author{Yu.Shitov for NEMO collaboration}

\address{JINR,\\
Dubna, Moscow region, \\
141980, Russia\\
E-mail: yuriy@nu.jinr.ru}

\maketitle

\abstracts{ 
The NEMO-3 detector installed in the Modane Underground Laboratory
(LSM, France) is running to search for neutrinoless double beta decay
($\beta\beta$0$\nu$) with an expected sensitivity for the effective
Majorana neutrino mass down to 0.1~eV. New preliminary limits (90\%
CL), $T_{1/2}(\beta\beta0\nu) > 1.0 \cdot 10^{23}$ y (right-handed
currents) and $T_{1/2}(\beta\beta0\nu) > 1.9 \cdot 10^{23}$ y (massive
mechanism) were established analyzing $^{100}$Mo data collected during
3800h of measurements in 2003. This is the world's best result for
$^{100}$Mo.  New preliminary half-lives of $\beta\beta$2$\nu$-mode for
$^{100}$Mo, $^{82}$Se, $^{116}$Cd, $^{150}$Nd are also reported as
well as the results of background studies. The future progress and
plans of the NEMO-3 project are discussed.}

\section{Introduction}
\label{intro}

Recent observations of neutrino oscillations reported by a number
of experiments\cite{SuperK} \cite{SNO} \cite{KamLand} have proved 
the existence of non-zero neutrino mass. This outstanding  progress
has triggered a renewed interest in neutrino physics. The search for
neutrinoless double beta decay ($\beta\beta$0$\nu$) together with 
new generation neutrino oscillation experiments will help solve the
neutrino mass puzzle. 
The discovery of this process would confirm the Majorana nature of
neutrinos as well as it will determine the absolute mass scale and the 
pattern of neutrino eigenstates. Moreover, the $\beta\beta$0$\nu$-search seems to
be now the only method which could give the answers on the question 
of neutrino mass nature and full lepton number non-conservation.

\section{The NEMO-3 detector}
\label{sec:experiment}

The main goal of the NEMO-3 experiment\cite{proposal} is to search
for neutrinoless double beta decay with a sensitivity 
to the half-life of up to $\mathrm{10^{25}}$
years. The NEMO-3 detector was built in the Fr\'ejus Underground
Laboratory(LSM) which has an overburden of 4800 m.w.e. to reduce 
the cosmic ray flux.

\begin{center} 
\begin{figure*}[ht] 

\centerline{\epsfxsize=10cm \epsfbox{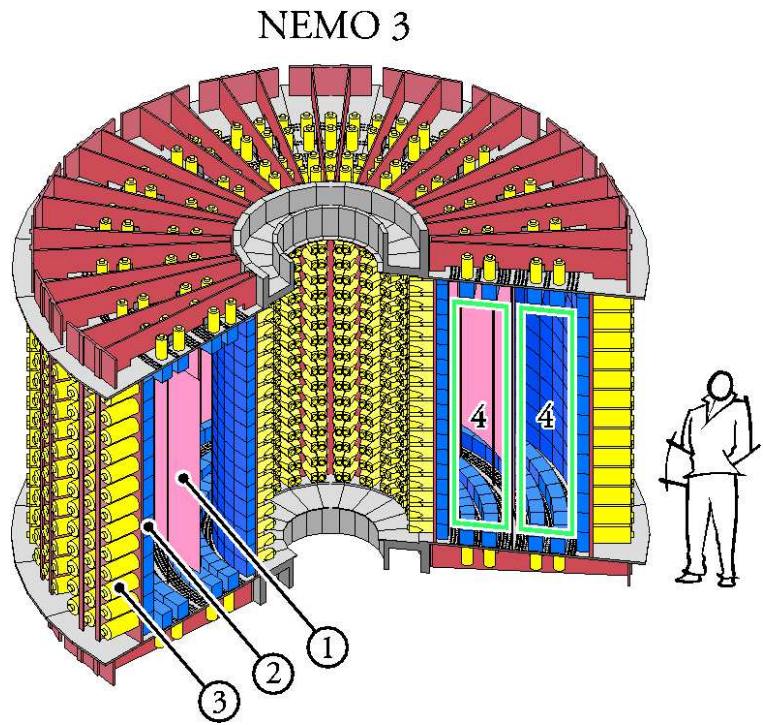}}

\caption{Schematic view of the NEMO-3 detector core without external
shielding: (1) source foil, (2) 1940 plastic scintillators coupled to
(3) low activity photomultiplier tubes, (4) tracking volume with 6180
drift cells operating in Geiger mode}
\label{fig:nemo3}
\end{figure*}
\end{center}

Details of the detector's technical design have been already reported
by the NEMO collaboration\cite{Ohsumi} and will be published
soon\cite{NEMO-design}. We will only briefly describe here 
the basic principles of the setup shown in Figure~\ref{fig:nemo3}.
The electrons emitted from the thin foils of a $\beta\beta$-source 
go through the
tracking chamber and hit scintillators of the calorimeter. A separate
registration of tracks and energies of both $\beta\beta$-electrons is
the main advantage of the so-called ''tracko-calo'' method exploited by
the NEMO-3 detector. It allows a real observation of $\beta\beta$-decay
measuring all its characteristics. Some important information, such as
the angular correlation between $\beta\beta$-electrons and single
$\beta$-spectra, can not be extracted from other types of
$\beta\beta$-measurements.

At the same time the background is tagged, monitored, and rejected by
the tracko-calo technique with a very high efficiency using several
mechanisms:
\begin{itemize}
\item Time-of-flight (TOF) analysis allows the rejection of external electrons 
crossing the tracking volume;
\item tracking information allows the rejection of events without common vertex;
\item A magnetic field of \~25 Gauss directed along the detector axis allows 
the rejection of electron-positron pairs created by external photons interacting 
in the source foil;
\item Explicit identification of electrons, gammas and delayed-alphas gives a
unique opportunity to monitor and reject different background events;
\end{itemize}

The main isotopes under study to search for $\beta\beta$0$\nu$-decay are
$\mathrm{^{100}}$Mo (6.9 kg) and $\mathrm{^{82}}$Se (0.9 kg).
$\mathrm{^{116}}$Cd (0.4 kg), $\mathrm{^{130}}$Te (0.5 kg),
$\mathrm{^{150}}$Nd (37 g), $\mathrm{^{96}}$Zr (9 g) and
$\mathrm{^{48}}$Ca (7 g) are used to study two-neutrino double beta
decay ($\beta\beta$2$\nu$). Other sources (0.6 kg of natural Tellurium
oxide and 0.6 kg of Copper) are installed to study the background.
Some of the sources have been purified to reduce their
contamination with $\mathrm{^{214}Bi}$ and $\mathrm{^{208}Tl}$, which are 
the main two
isotopes with a large $\mathrm{Q_{\beta}}$-value, that can potentially
contribute to the background in the $\beta\beta$0$\nu$ energy window.

The design allows radioactive sources to be introduced 
inside the NEMO-3 detector for calibration
purposes.  $\mathrm{^{207}}$Bi and $\mathrm{^{90}}$Sr are used for
the absolute energy calibration ($\sigma/E$ \~6$\%$ for 1~MeV
electrons) while the time calibration is performed
with $\mathrm{^{60}}$Co, $\mathrm{^{207}}$Bi, and neutron sources
placed outside the detector. The neutron runs are also used to check
the performance of the tracking chamber using high energy electrons crossing
the tracking volume. The transverse and longitudinal vertex resolutions
obtained at 1 MeV are 0.2 and 0.7 cm, respectively. A laser-based
light injection system is
used for a daily survey of the calorimeter (relative energy calibration).

The detector core is surrounded by multi-layer shielding, which
includes a solenoid's copper coil, 18 cm of iron to suppress gamma flux,
and finally 35 cm of water in tanks or wood against neutrons.  All
components of the detector have been tested to have extra low
radioactivity.

First tests confirmed that the performance of the NEMO-3
detector corresponds to that claimed in the proposal\cite{proposal}.

\section{Results of analysis of 2003 data}
\label{sec:physics}

3800 h of data collected between February and October 2003 have been
analyzed.

\subsection{Background studies}

A great advantage of the NEMO-3 detector is its ability to measure
various backgrounds by detecting electrons,
gammas and delayed alphas in any combination with different
multiplicities.

Our studies were specially concentrated on the backgrounds, which
could contribute to the $\beta\beta$0$\nu$-mode. In addition to unavoidable
background from $\beta\beta$2$\nu$-tail these are $\mathrm{^{208}Tl}$
(located in the source foil) and $\mathrm{^{214}Bi}$ (on the foil's surface
or close to it in the gas) originated from the $\mathrm{^{222}Rn}$ decay
chain.

A $\beta\beta$-like event may come from a ($\beta$,$\gamma$) decay of
$\mathrm{^{214}Bi}$ or $\mathrm{^{208}Tl}$ followed by the production of
the second electron either by the photon via the Compton effect, or by the
electron via the M\"oller scattering or alternatively as a result of
a conversion electron emitted instead of the photon\cite{Vasiliev}.

$\beta\gamma$, $\beta\gamma\gamma$, and $\beta\gamma\gamma\gamma$
channels have been studied to estimate the $\mathrm{^{208}Tl}$
contamination in the source foils. The TOF and high-energy cuts for
electrons and photon(s) were used to isolate
$\mathrm{^{208}Tl}$-events from other backgrounds.  The
$\mathrm{^{208}Tl}$ activities in composite and metallic foils of
$\mathrm{^{100}Mo}$ were found to be 140 $\pm$ 40 and 50 $\pm$ 40 $\mu
Bq/kg$, respectively. This is in good agreement with the limit
obtained with a Ge detector ($<$100 $\mu Bq/kg$). The accuracy of
the $\mathrm{^{208}Tl}$ activity estimate will be improved
in the future with higher statistics.

$\mathrm{^{214}Bi}$ is a daughter isotope of the $\mathrm{^{222}Rn}$
decay chain.  Radon penetrates into the NEMO-3 inner volume from the
laboratory air.  The $\mathrm{^{214}Bi}$ contamination is estimated
from $\beta\alpha$-events with one electron track followed by an alpha
track (delayed Geiger hits) originated from the same vertex. The time
distribution of the delayed alphas corresponds to the decay constant
of $\mathrm{^{214}Po}$ (T$_{1/2}$=164$\mu$s) confirming the efficient
selection of $\mathrm{^{214}Bi}$-events.  The $\mathrm{^{222}Rn}$
activity reconstructed from the $\beta\alpha$-channel is at the level
of 30 $\mathrm{mBq/m^3}$ and oscillates with time. This value is in
agreement with the results of independent measurements of the NEMO-3
tracking chamber gas made by a special radon detector at the
detector's gas outlet. A daily survey of $\mathrm{^{222}Rn}$ activity
inside NEMO-3 is performed now using the method described. Higher
statistics will allow us to determine relative activities of
$\mathrm{^{214}Bi}$ in the foil, Geiger wires and gas volume.

A few $\beta\beta$0$\nu$-like events per year are expected from
the $\mathrm{^{214}Bi}$ decay given the current contamination of radon inside
NEMO-3, which is by far more than other background contributions. In
order to reach the expected sensitivity for $\beta\beta$0$\nu$, the radon level 
in NEMO-3 must be reduced by a factor of 10$-$100. 
To achieve this the detector was
surrounded by an anti-radon tent. A radon free air factory analogous to one
used by SuperKamiokande is currently being built in LSM. It is expected that
the anti-radon tent will be purged with a purified air stream 
starting from this summer.

\subsection{Two-neutrino double beta decay}
\label{sec:2}

The $\beta\beta$-event selection criterion requires two electron
tracks originated from a common vertex in the source and associated
with two scintillator hits. To suppress
the $\mathrm{^{214}Bi}$-contribution, events with delayed hits ($\alpha$
particle) near the vertex are rejected.

\begin{figure}
\resizebox{0.47\textwidth}{0.3\textheight}{\includegraphics{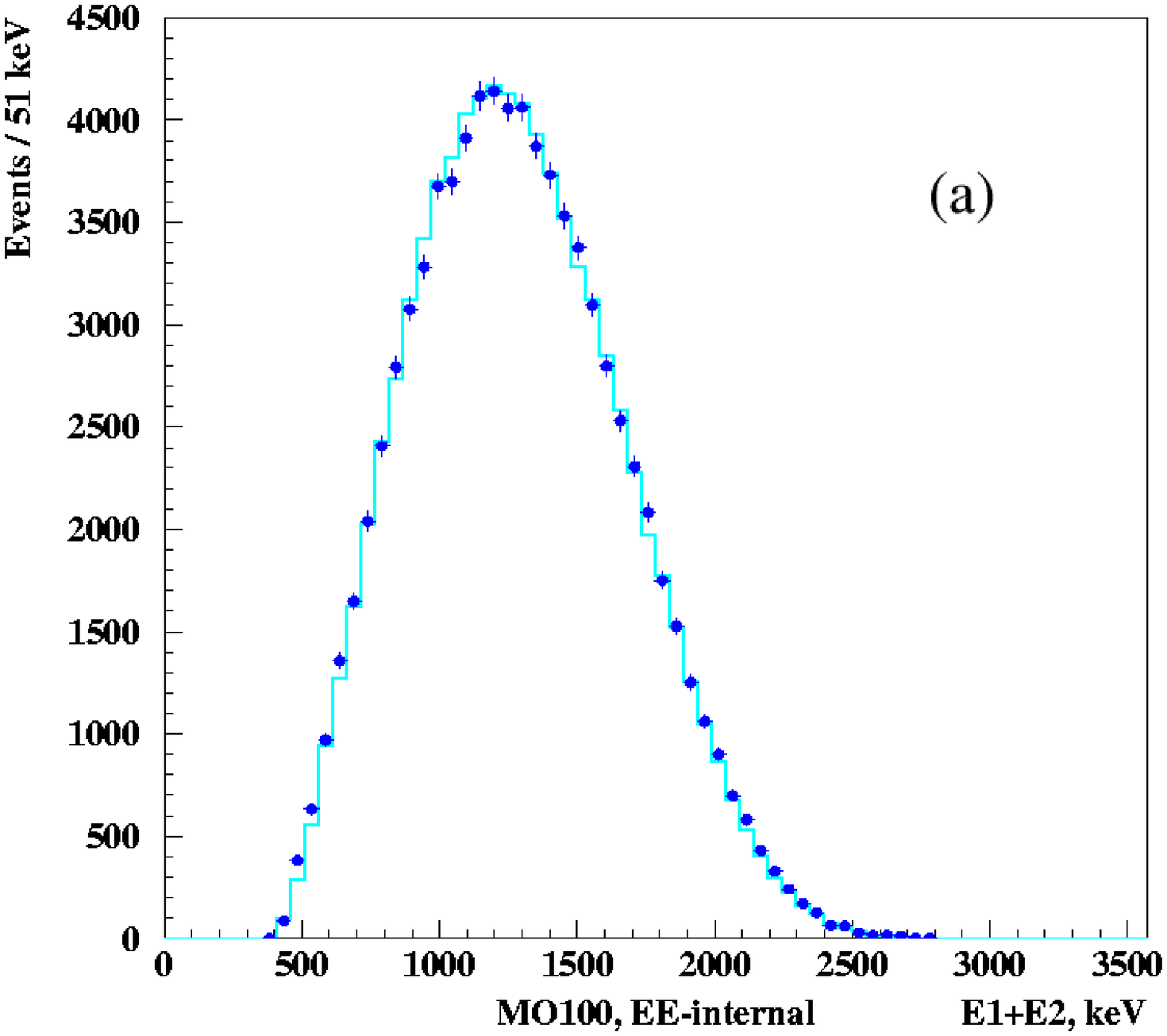}}
\resizebox{0.47\textwidth}{0.3\textheight}{\includegraphics{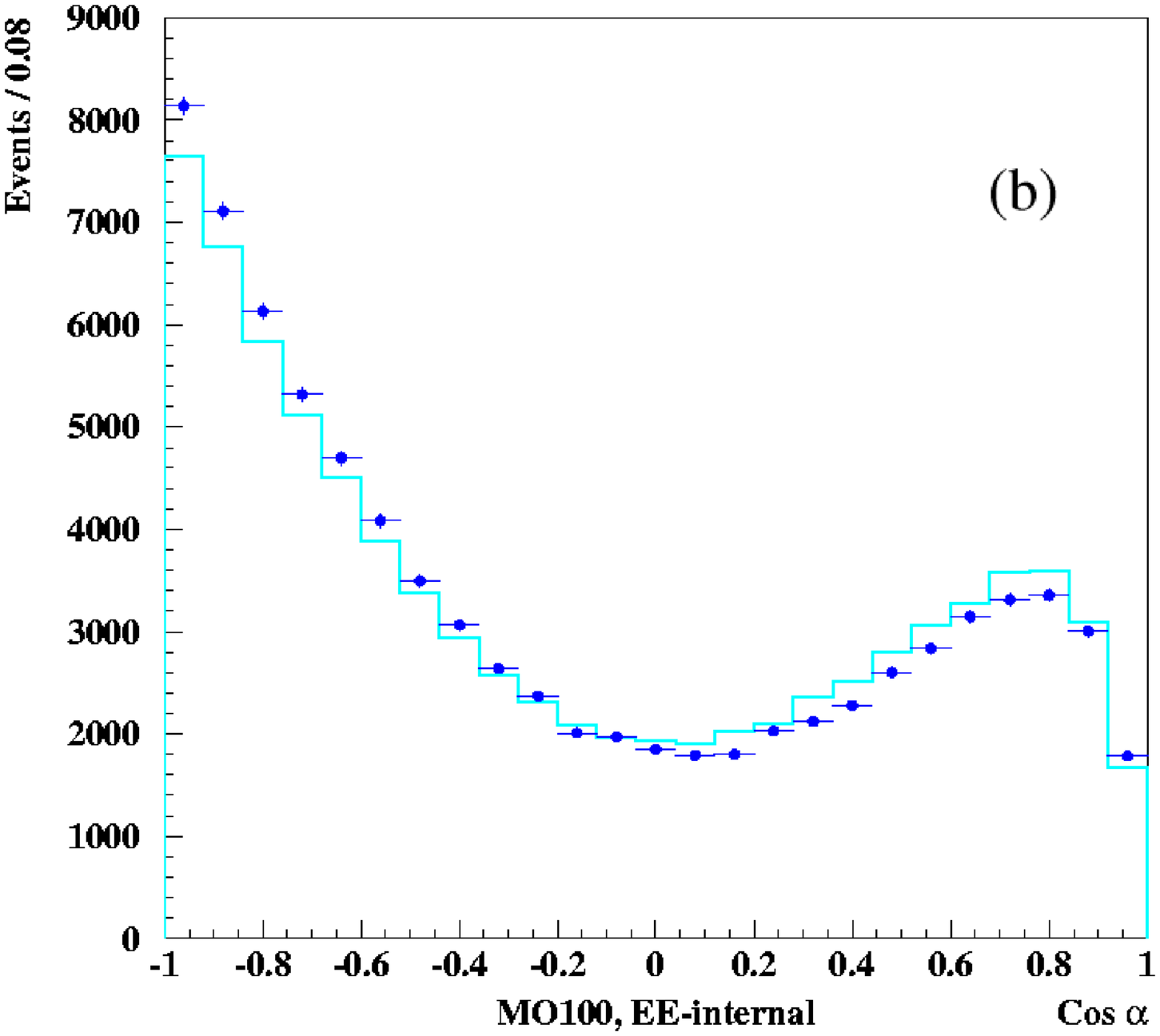}}
\caption{Distribution of total energy of electrons (a) and cosine of 
angle between two electrons (b) for $\mathrm{^{100}Mo}$
$\beta\beta$-events after 3800~h of data taking. The points
represent data with background subtracted while the
histograms show 2$\beta$2$\nu$-simulations.}
\label{mo}
\end{figure}

80000 events were selected in the $\mathrm{^{100}Mo}$ sources with a
signal-to-background (S/B) ratio of 40. This statistics is seven times
the one collected by the NEMO-2 detector\cite{n2mo100_1}.  The energy sum of
the two electrons as well as the angular distribution between
them are shown in Figure~\ref{mo}.

The preliminary value of the $\beta\beta$2$\nu$ decay half-life for
$\mathrm{^{100}Mo}$ is:

$$\mathrm{T_{1/2}^{\beta\beta2\nu} = 7.3 \pm 0.025 (stat) \pm 0.7
(syst) 10^{18} y.}$$

Due to uncertainties in the detector efficiency, a conservative 10$\%$
systematic error has been used.

\begin{figure}
\resizebox{0.9\textwidth}{0.25\textheight}{\includegraphics{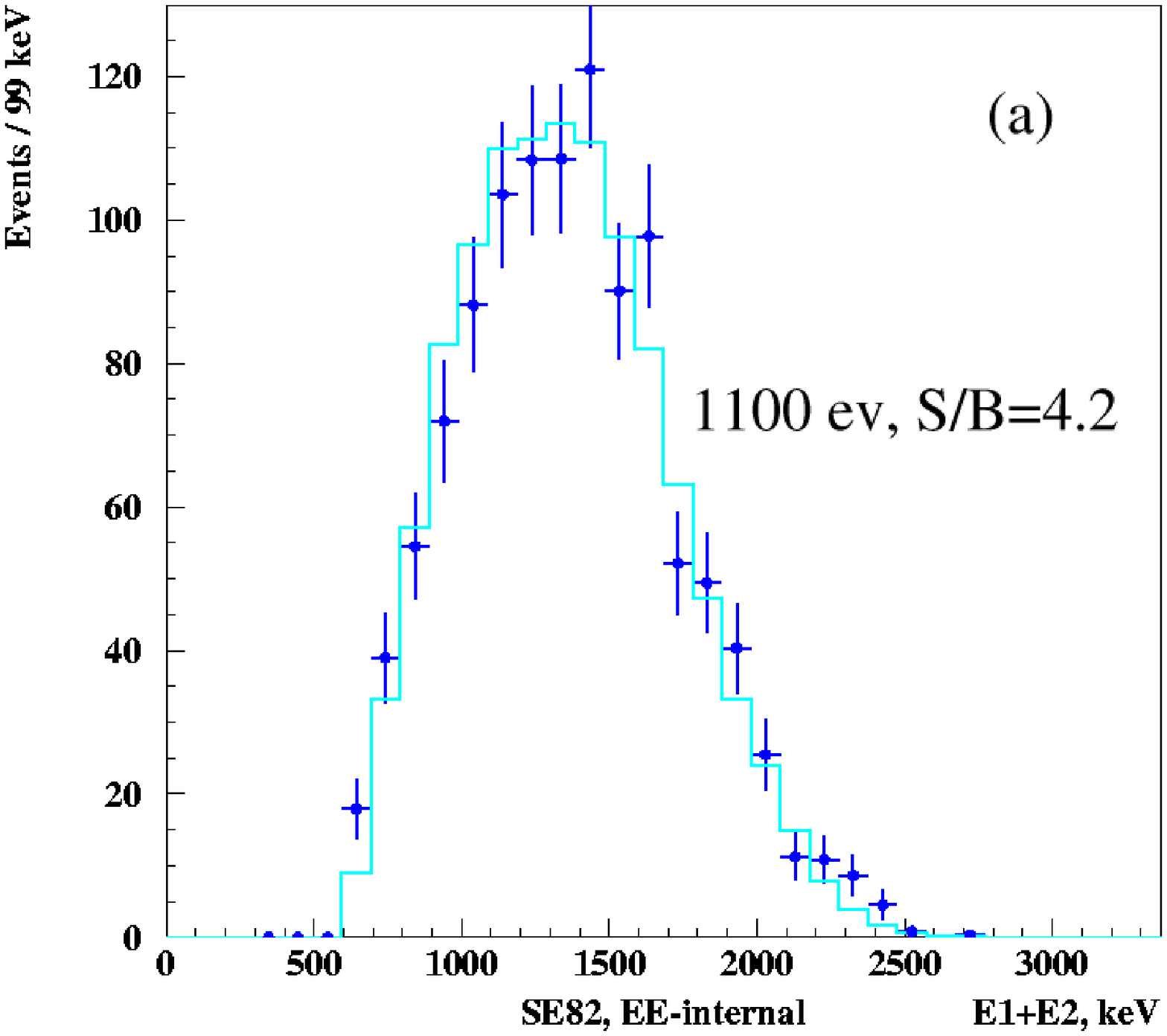}}
\resizebox{0.47\textwidth}{0.25\textheight}{\includegraphics{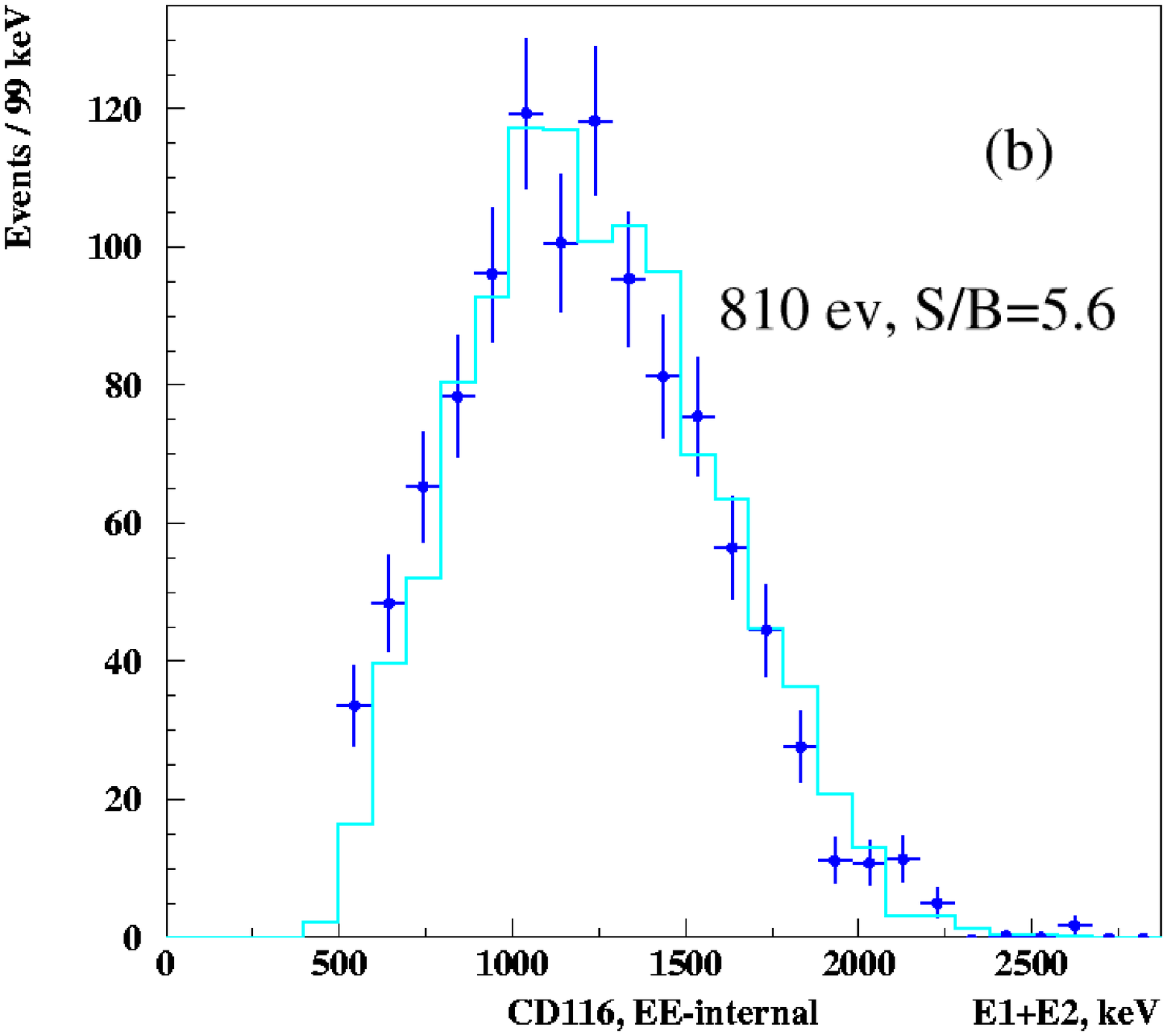}}
\resizebox{0.47\textwidth}{0.25\textheight}{\includegraphics{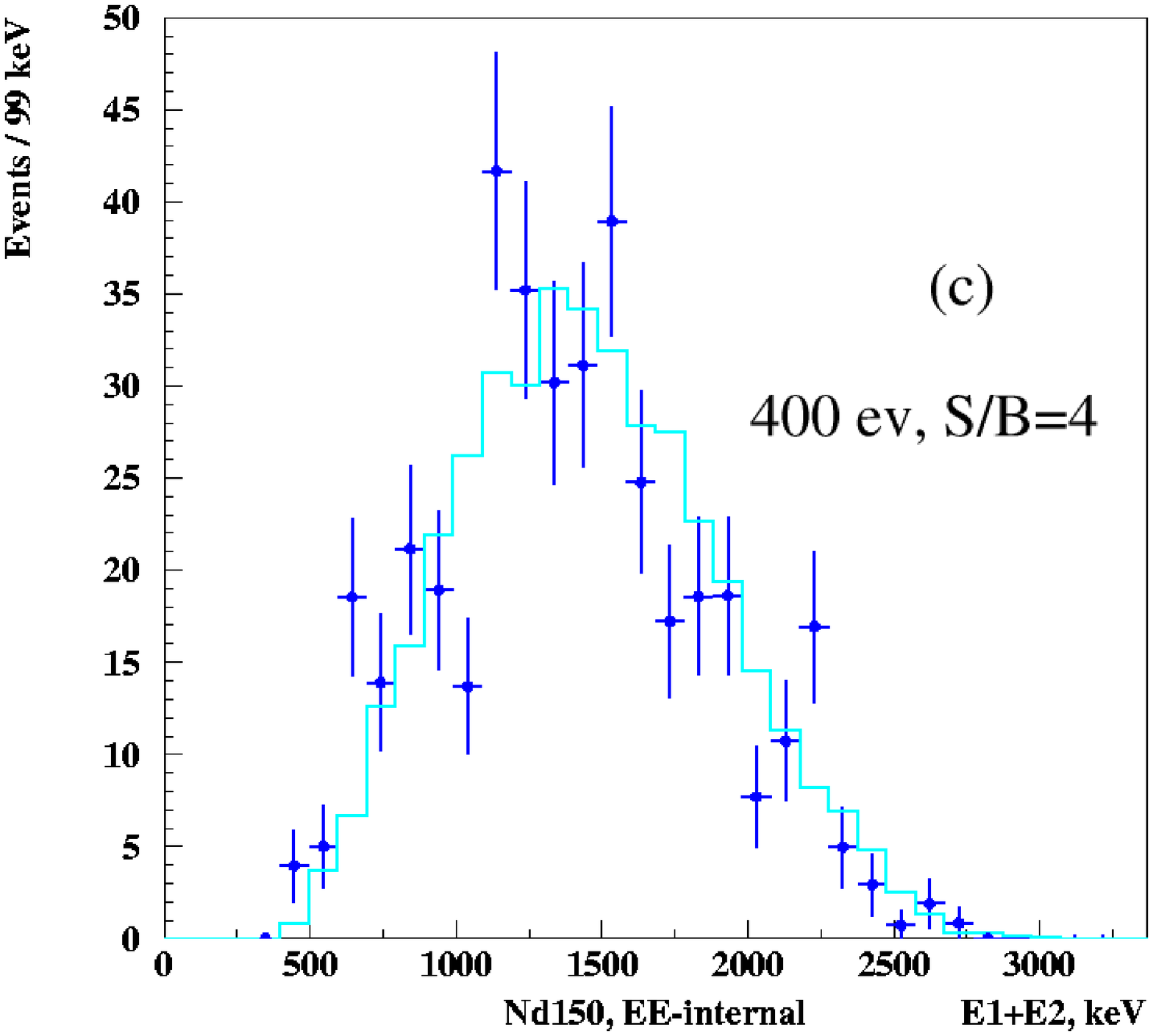}}
\caption{Distribution of total energy of $\beta\beta$-electrons for 
$\mathrm{^{82}Se}$ (a), $\mathrm{^{116}Cd}$ (b), and
$\mathrm{^{150}Nd}$ (c), after 3800~h of data taking. The points
represent data with background subtracted while the
histograms show 2$\beta$2$\nu$-simulations.}
\label{other}
\end{figure}

Other isotopes were analyzed in the same way. The spectra of the electron energy
sum are shown in Figure~\ref{other}, while the preliminary
values of the $\beta\beta$2$\nu$ half-lives are given below:

\begin{tabular}{ccll}
$\mathrm{^{82}Se}$ &:&
  $\mathrm{T_{1/2}^{\beta\beta2\nu} 
   = 9.5 \pm 0.3 (stat) \pm 0.9 (syst) 10^{19} y.}$\\
$\mathrm{^{116}Cd}$ &:&
  $\mathrm{T_{1/2}^{\beta\beta2\nu}  
   = 2.7 \pm 0.1 (stat) \pm 0.3 (syst) 10^{19} y.}$\\
$\mathrm{^{150}Nd}$ &:&
  $\mathrm{T_{1/2}^{\beta\beta2\nu} 
= 7.5 \pm 0.3 (stat) \pm 0.7 (syst) 10^{18} y.}$\\
\end{tabular}

\subsection{Neutrinoless double beta decay}

The $\beta\beta$0$\nu$ decay can be produced by V-A or V+A
currents. These two modes can be distinguished by studying the angular
distribution of the emitted electrons. An analysis with a likelihood
method which took into account the minimal single electron energy, total
energy and the angular distribution of $\beta\beta$-electrons has been
performed.

The limits (90~\%~CL) obtained for $\mathrm{^{100}Mo}$ are:

\begin{tabular}{lcll}
V-A &:&
  $\mathrm{T_{1/2}^{\beta\beta0\nu} > 1.9 \; \cdot \; 10^{23} y.}$\\
V+A &:&
  $\mathrm{T_{1/2}^{\beta\beta0\nu} > 9.8 \; \cdot \; 10^{22} y.}$\\
\end{tabular}

These limits improve the best current results obtained for
$\mathrm{^{100}Mo}$\cite{Ejiri}.

\section{Discussion and conclusion}
\label{sec:conclusion}

The first data collected by NEMO-3 has shown the detector to perform 
within the original specifications.
Since February 2003 the data has been taken routinely. 
After a few months of data taking the NEMO-3 experiment has already
produced the world's  best limit for the $\beta\beta$0$\nu$-mode of 
$\mathrm{^{100}Mo}$. After 5 years of measurements we expect 
to reach a  sensitivity to the effective Majorana neutrino mass of
0.1~eV.

It should be emphasized, that 0.1~eV is the first milestone on the current
road map of the $\beta\beta$0$\nu$-search and a short-term (5~years)
goal for current $\beta\beta$-experiments. If $\beta\beta$0$\nu$-decay
is not found at this level, the degenerated hierarchy
pattern can be excluded for Majorana neutrinos.

According to current theoretical assumptions the second milestone is
the sensitivity at the level of 0.02-0.04~eV corresponded to the
inverted hierarchy of neutrino eigenstates. This is a longer-term
($\geq$10$-$20 years) goal of many ambitious projects (EXO, MOON,
MAJORANA, etc.) with hundreds, and possibly tons, of kilograms of
enriched $\beta\beta$-isotopes. In the end of 2003 the NEMO
collaboration launched a R\&D program to study the feasibility of a
NEMO-like detector with $\sim$100 kg of isotope and improved energy
resolution in order to reach this sensitivity (SUPERNEMO).  Such a
project would represent the culmination of two decades of development
which has provided three previous NEMO experiments.  Because of the
modest R\&D needed and very well known technology, this future project
can potentially have a significant head start on the competition.

\section*{Acknowledgments}
\label{sec:acknowledgements}

The authors would like to thank the Fr\'ejus Underground Laboratory
staff for their technical assistance in building and running the
experiment.

This work was partially supported by INTAS Grant No~03051-3431.

\end{document}